\documentclass[10pt,aps,pra,showpacs]{revtex4-1}
\usepackage{amsmath,amssymb,amsfonts,latexsym,dcolumn}
\usepackage[dvipsnames]{xcolor}
\usepackage{graphicx,psfrag,amsmath,amssymb,amsfonts,latexsym,color,dcolumn,bm, subfig}
\usepackage{hyperref}
\usepackage[latin1]{inputenc}
\usepackage[T1]{fontenc} 
\usepackage[english]{babel}

\begin{document}
\title{Motion induced radiation and quantum friction for a moving atom}

\author{M. Bel\' en Far\'\i as $^{1,2}$ \footnote{mbelfarias@df.uba.ar}}
\author{C. D. Fosco $^3$ \footnote{fosco@cab.cnea.gov.ar}}
\author{Fernando C. Lombardo$^1$ \footnote{lombardo@df.uba.ar}}
\author{Francisco D. Mazzitelli$^3$ \footnote{fdmazzi@cab.cnea.gov.ar}}

\affiliation{$^1$ Departamento de F\'\i sica {\it Juan Jos\'e
 Giambiagi}, FCEyN UBA, Facultad de Ciencias Exactas y Naturales,
 Ciudad Universitaria, Pabell\' on I, 1428 Buenos Aires, Argentina }
\affiliation{$^2$ University of Luxembourg, Physics and Materials Science
Research Unit, Avenue de la Fraïncerie 162a, L-1511, Luxembourg, Luxembourg}
\affiliation{$^2$ Centro At\'omico Bariloche and Instituto Balseiro,
Comisi\'on Nacional de Energ\'\i a At\'omica, 
R8402AGP Bariloche, Argentina}
\date{today}

\begin{abstract}
We study quantum dissipative effects that result from the non-relativistic
	motion of an atom, coupled to a quantum real scalar field, in the
	presence of a static imperfect mirror.  Our study consists of two
	parts: in the first, we consider accelerated motion in free space,
	namely, switching off the coupling to the mirror. This results in
	motion induced radiation, which we quantify  via the vacuum
	persistence amplitude. In the model we use, the atom is described
	by a quantum harmonic oscillator (QHO). We show that its natural frequency poses a threshold
	which separates different regimes, involving or not the internal
	excitation of the oscillator, with the ulterior emission of a
	photon.  At higher orders in the coupling to the field, pairs of
	photons may be created by virtue of the Dynamical Casimir Effect
	(DCE).  
\noindent In the second part, we switch on the coupling to the mirror,
	which we describe by localized microscopic degrees of freedom. 
	We show that this leads to the existence of quantum contactless friction as well as to corrections to 
	the free space emission considered
	in the first part.  The latter are similar to the effect of a
	dielectric on the spontaneous emission of an excited atom.  We have
	found that, when the atom is accelerated and close to the plate, it
	is crucial to take into account the losses in the dielectric in
	order to obtain finite results for the vacuum persistence
	amplitude.
\end{abstract}
\maketitle 


\section{Introduction}\label{sec:intro}

Many interesting physical phenomena arise when quantum systems are
subjected to the influence of external time-dependent conditions.  For
instance, accelerated neutral objects may radiate photons, even in the
absence of permanent dipole moments. This is the so called {\em motion
induced radiation} or Dynamical Casimir effect (DCE)\cite{reviews}. On
the other hand, neutral objects moving sidewise with constant relative speed
may influence each
other by a frictional force proportional to a power of the velocity
(quantum friction)\cite{qfriction}.

In this work, we study quantum dissipative effects which are due to the motion 
of an atom coupled to a vacuum real scalar field. We consider 
the cases of an isolated atom and an atom
in the presence of a (planar) plate. The latter will be assumed to 
behave as an `imperfect' mirror regarding the reflection/transmission properties 
it manifests, under the propagation of vacuum-field waves. 

Our description of the microscopic degrees of freedom will be similar 
for both the plane and the atom. Indeed, in both cases, they 
will be assumed to be modes linearly coupled to the vacuum field, 
and to have a harmonic-oscillator like action, with an intrinsic frequency 
parameter. We assume the plate to be homogeneous, so that the frequency will be 
one and the same for all the points on the plate. This is 
essentially the model considered in~\cite{Farias:2014wca} in which
we analyzed quantum friction, except that here we also include a damping
parameter, to account for losses in the dielectric.  For the point-like
particle, on the other hand, we use a single harmonic oscillator, with a 
linear coupling to the vacuum field. In Ref.~\cite{ludmila} thermal corrections were 
also considered.  

We use a model based on the assumptions above to derive the vacuum
persistence amplitude as a functional of the trajectory of the particle,
for different kinds of motion.  Our goal is to explore the  internal
excitation process of the atom with emission of a photon, pairs creation of
photons due to DCE, the appearance of quantum contactless friction, and the
corrections to free emission which are due to the presence of the mirror. 

Regarding related works,  the relevance of the internal degrees of freedom
of the plate in the context of optomechanics has been analyzed and reviewed
in Ref.~\cite{HuMof}.  In Ref.\cite{PAMN}, the radiation produced by an
atom moving non relativistically in free space has been studied in detail
(see also Ref.\cite{Law}). It was shown there that, when the atom
oscillates with  a mechanical frequency smaller than the internal
excitation energy, the radiation produced, that consists of photon pairs,
can be considered as a  microscopic counterpart of the DCE. In the opposite
regime, the atom becomes mechanically excited, and then emits single
photons returning to its ground state.  Accelerated harmonic oscillators
have also been considered in the context of the Unruh effect, as toy models
for particle detectors~\cite{Unruheffect}. 

We generalize here previous analyses, to account for the presence of a
plate, treating in a unified fashion photon emission and quantum friction.
The possibility of enhancing the quantum friction forces by considering
arbitrary angles between the atom's direction of motion and the surface has
been discussed in Ref.\cite{DalvitBelen}. It has also been shown that the
presence of a plate may influence the fringe visibility  in an atomic
interference experiment (see Refs.\cite{Villanueva,Ccappa}).  A molecule
moving with constant speed over a dielectric with periodic grating can show
parametric self-induced excitation and, in turn, it can produce a
detectable radiation \cite{Capasso}.  Note that this situation can be
mimicked by the superposition of  constant velocity and oscillatory motions
over a flat surface. Although different, this phenomenon reminds the
classical Smith-Purcell radiation for charged objects moving with constant
velocity over a periodic grating, and its  eventual influence on a
double-slit experiment with electrons \cite{quantumSmithPurcell}.  The
problem of moving atoms near a plate is also relevant when discussing
dynamical corrections to the Casimir-Polder interaction
\cite{CasimirPolder}.

This paper is organized as follows: in Section~\ref{sec:thesys}, we
introduce the model for a particle in free space and define its effective
action.  Then in Section~\ref{sec:no mirror} we evaluate that effective
action perturbatively in the coupling between the atom and field. To the
leading order in a weak-coupling expansion, there is a threshold for the
imaginary part of the effective action, associated to the internal
excitation of the atom before radiation emission. The next-to-leading order (NTLO) shows the
combination of this effect and the usual Dynamical Casimir effect, that
does not involve such excitation.  In Section \ref{sec:model mirror} we
introduce the model for the imperfect mirror, considering quantum harmonic
oscillators as microscopic degrees of freedom coupled to an environment as
a source of internal dissipation.  In Section~\ref{sec:mirror} we evaluate
the vacuum persistence amplitude for the case of an atom moving near the
plate, up to first order in both couplings (atom-field and mirror-field). 
We apply the general expressions for the imaginary part of the effective
action to the calculation of dissipative effects, for qualitatively
different particle paths, and look for effects of quantum friction and
motion induced radiation separately.  We will see that, due to resonant
effects,  the internal dissipation of the mirror is crucial to obtain
finite results.  We present our conclusions in  Section~\ref{sec:conc}.  
\section{The system and its effective action: isolated particle}\label{sec:thesys}

Throughout this paper, we consider the non-relativistic motion of a point
particle in three spatial dimensions, with a trajectory described by $t \to {\mathbf r}(t)
\in {\mathbb R}^3$, with $|\dot{\mathbf r}(t)| < 1$ (we use natural
units, such that $c = 1$ and $\hbar =1$).

We then introduce the in-out effective action,
$\Gamma[{\mathbf r}(t)]$, a functional of the particle's trajectory,
which is defined by means of the expression:
\begin{equation}\label{eq:defgrt}
	e^{ i \Gamma[{\mathbf r}(t)]} \;\equiv\; 
	\frac{\int {\mathcal D}\phi \; e^{i {\mathcal S}(\phi)}}{\int {\mathcal
D}\phi \; e^{i {\mathcal S}_0(\phi)}}
\,=\, 
\frac{\int {\mathcal D}\phi \; e^{i {\mathcal S}_I(\phi)} \;
e^{i {\mathcal S}_0(\phi)}}{\int {\mathcal
D}\phi \; e^{i {\mathcal S}_0(\phi)}}
\,\equiv \, 
\langle \, e^{i {\mathcal S}_I(\phi)} \, \rangle_0 \;, 
\end{equation}
where the functional integrals are over $\phi(x)$, a vacuum real scalar
field in $3+1$ dimensions, equipped
with an action ${\mathcal S}$, which consists of two terms:
\begin{equation}\label{eq:defs}
	{\mathcal S}(\phi) \;=\; {\mathcal S}_0(\phi) \,+\, {\mathcal
	S}_I^{(p)}(\phi) \;.
\end{equation}
${\mathcal S}_0$, denotes the part of the action which describes its
free propagation:
\begin{equation}
{\mathcal S}_0(\phi) \;=\; \frac{1}{2} \,\int d^4x \left[ \partial_\mu
	\phi (x) \partial^\mu \phi (x) + i \epsilon \phi^2(x) \right] \;,
	\;\; x = (x^0,x^1,x^2,x^3) \;,
\end{equation}
while ${\mathcal S}_I^{(p)}$ represents the coupling of the scalar field to the
particle. In the kind
of model that we consider here, it is assumed to be
quadratic, namely:
\begin{equation}
	{\mathcal S}_I^{(p)}(\phi) \;=\; -\frac{1}{2} \int_{x,y} \phi (x)
	V_{p}(x,y)
	\phi(y) \;,
\end{equation}
where we have introduced a shorthand notation for the integration over
space-time points. The kernel, $V_p$  is a `potential' resulting from the
integration of microscopic degrees of freedom. It can be regarded as a
symmetric function of $x$ and $y$. In what follows, we consider its form in more detail. 

Assuming a single degree of freedom which corresponds  bosonic oscillator,
endowed with a coordinate $q$, living on the particle's
internal space, that potential $V_p$ stems from the functional integral over $q$:
\begin{equation}\label{eq:gq}
e^{- \frac{i}{2} \int_{x,y} \phi(x) V_p(x,y) \phi(y) } 
 \;=\; 
\int {\mathcal D}q \; e^{i {\mathcal S}_p(q,\phi;{\mathbf r})} \;,
\end{equation}
where the particle's action, ${\mathcal S}_p$, is given by:
\begin{align}\label{eq:defspo}
{\mathcal S}_p(q,\phi;{\mathbf r}) &=\;{\mathcal S}_p^{(0)}(q) 
\,+\,{\mathcal S}_p^{int}(q,\phi;{\mathbf r}) \;, 
\nonumber\\
	{\mathcal S}^{(0)}_p(q) &=\;\frac{1}{2}\int dt \, 
	\big( \dot{q}^2 - ( \Omega_p^2 - i \epsilon) q^2 \big) \;,
\nonumber\\
{\mathcal S}_p^{int}(q,\phi;{\mathbf r}) &=\; g \,
	\int dt \,  q(t) 
\; \phi(t,{\mathbf r}(t)) \;.
\end{align}
Here, $\Omega_p$ is the harmonic oscillator frequency, and $g$ determines
the coupling between the oscillator and the real scalar field. Note that
$g$ has the dimensions of  $[{\rm mass}]^{1/2}$. 

We see that:
\begin{equation}
V_p(x,y) \;=\;  g^2 \, 
\delta\big({\mathbf x} - {\mathbf r}(x^0)\big) 
\,
\Delta_p(x^0-y^0)	
\,
\delta\big({\mathbf y} - {\mathbf r}(y^0)\big) 
\end{equation}
where:
\begin{equation}\label{eq:defdp}
\Delta_p(x^0 - y^0) \;=\; \int  \frac{d\nu}{2\pi} \, e^{-i \nu
	(t-t')} \, \widetilde{\Delta}_p(\nu) \;,\;\;\;
	\widetilde{\Delta}_p(\nu)  \,\equiv \, \frac{1}{\nu^2 - \Omega_p^2
	+ i \epsilon} \;.
\end{equation}

We could proceed in the alternative way, and integrate first the scalar field in order to obtain an effective action
for the harmonic oscillator. Although we will not follow this approach here, for later use we note that
such integration gives, when ${\mathbf r}=0$, 
\begin{equation}
{\mathcal S}_p^{eff}(q) = {\mathcal S}_p^{(0)}(q)-\frac{g^2}{2}\int dt\, dt' q(t)G_0(t-t')q(t')\, ,
\end{equation}
where $G_0(t-t')$ is the Feynman propagator for the scalar field evaluated at coincident spatial points
\begin{equation}
G_0(t-t')=\int \frac{d^4 p}{(2\pi)^4}\frac{e^{-i p_0(t-t')}}{p_0^2 -{\mathbf p}^2+i\epsilon}
\end{equation}
The integral over the spatial momentum is linearly divergent, and the propagator becomes proportional 
to $\Lambda \delta(t-t')$, where $\Lambda$ is $3-$momentum cutoff. This divergence produces a 
shift $\delta\Omega$ in the natural frequency of the oscillator
\begin{equation}\label{eq:renOmega}
\Omega_p + \delta\Omega = \Omega_p^{(ren)}\, ,\,\,  \delta\Omega=-\frac {g^2}{4\pi^2}\frac{\Lambda}{\Omega_p^{(ren)}}\, .
\end{equation}
The divergence is of course a consequence of considering point-like interactions with the field.

The effective action, $\Gamma_p[{\mathbf r}(t)]$, is a functional of the trajectory and is given by:
\begin{equation}\label{eq:gp}
	e^{i\Gamma_p[{\mathbf r}(t)]} \;=\; \left\langle   e^{-\frac{i}{2} \int_{x,y} \phi(x)
	V_p(x,y) \phi(y)} \right\rangle_0 \;,
\end{equation}
where the average is taken with the free field action. The imaginary part of the effective action has the information of the dissipative effects due to the coupling of the 
moving harmonic oscillator and the field.

\section{Accelerated oscillator in free space}\label{sec:no mirror}

A perturbative expansion of $\Gamma_p[{\mathbf r}(t)]$ in powers of $V_p$
will produce a series of terms: \mbox{$\Gamma_p \,=\, \Gamma^{(1)}_p +
\Gamma^{(2)}_p + \ldots$}, where the index denotes the order in $V_p$. We
will consider just the first two terms in what follows, which already give
non-trivial results. The first-order term is given by:
\begin{equation}
\Gamma^{(1)}_p  = - \frac{1}{2} \, \int_{x,y}  V_p(x,y)  \big\langle
\phi(x) \phi(y)\big\rangle_0 \;,
\end{equation}
where $\langle \phi(x) \phi(y) \rangle_0 \equiv G_0(x,y)$, is the
Feynman propagator:
\begin{align}
G_0(x,y) &=\; \int \frac{d^4p}{(2\pi)^4} \, e^{-i p^0 (x^0-y^0)+i {\mathbf
p}\cdot ({\mathbf x} - {\mathbf y})} \, \widetilde{G}_0(p)
\;,\nonumber\\
\widetilde{G}_0(p) &=\; \frac{i}{(p^0)^2 - {\mathbf p}^2 + i \epsilon} \;,
\end{align}
while the second-order one, $\Gamma^{(2)}_p$, becomes:
\begin{equation}
\Gamma^{(2)}_p \;=\; \frac{i}{4} \, \int_{x,y,x',y'}
\,  V_p (x,y) \,  V_p (x',y') \, G_0(x,x') \, G_0(y,y') \;.
\end{equation}

Let us first evaluate $\Gamma^{(1)}_p$. 
Introducing the explicit forms of $V_p$ and $G_0$, we see that:
\begin{equation}
\Gamma^{(1)}_p \;=\; - \frac{g^2}{2} \,\int dx^0 \, \int dy^0 \;
\Delta_p(x^0-y^0) \; 
G_0(x^0-y^0,{\mathbf r}(x^0)- {\mathbf r}(y^0)) \;, 
\end{equation}
and, in terms of the respective Fourier transforms,
\begin{align}
\Gamma^{(1)}_p \;=\; - i \frac{g^2}{2}&\int \frac{d\nu}{2\pi} \, 
\int \frac{dp^0}{2\pi} \,\int \frac{d^3p}{(2\pi)^3}\, 
\int dx^0 \, \int dy^0 \,\Big[ \widetilde{\Delta}_p(\nu) \nonumber\\
&\times \frac{e^{-i (\nu + p^0) (x^0 - y^0)+ i {\mathbf p} \cdot ( {\mathbf r}(x^0)
- {\mathbf r}(y^0))}}{(p^0)^2 - {\mathbf p}^2 + i \epsilon} \Big]
\;.
\end{align}

Performing the shift $\nu \to \nu - p^0$,
\begin{equation}
\Gamma^{(1)}_p =\; \frac{1}{2}\,\int \frac{d\nu}{2\pi} \, 
\int \frac{d^3p}{(2\pi)^3}\, f(-{\mathbf p}, -\nu)
f( {\mathbf p}, \nu) \;
\Pi(\nu,{\mathbf p},\Omega_p) \;,
\end{equation}
\begin{equation}\label{eq:defpi1p}
\Pi(\nu,{\mathbf p},\Omega_p) \;\equiv\; - i g^2 \, 
\int \frac{dp^0}{2\pi} \; \frac{1}{(p^0 - \nu)^2 - \Omega_p^2 + i
\epsilon} \,\frac{1}{(p^0)^2 - {\mathbf p}^2 + i \epsilon} \;,
\end{equation}
where we have introduced:
\begin{equation}
f({\mathbf p},\nu) \,=\,
\int dt \,  e^{-i {\mathbf p}\cdot {\mathbf r}(t)} \;
e^{i\nu t} \;.
\end{equation}
After some algebra, and introducing a Feynman parameter $\alpha$,
($p \equiv |{\mathbf p}|$)
\begin{align}\label{eq:defpi}
	\Pi(\nu, p,\Omega_p) &=\; \frac{g^2}{4} \; \int_0^1 \, d\alpha \,
	\frac{1}{[D(\alpha, \nu, p)]^{3/2}} \nonumber\\ 
	D(\alpha, \nu, p) &\equiv\; \alpha \, \Omega_p^2 + (1-\alpha) \, p^2 -
	\alpha (1-\alpha) \, \nu^2  - i \epsilon \;. 
\end{align}

Therefore, 
\begin{equation}
{\rm Im} [ \Gamma^{(1)}_p]\;=\; \frac{1}{2} \,
	\int \frac{d\nu}{2\pi} \int \frac{d^3p}{(2\pi)^3} \;
	\big| f({\mathbf p}, \nu) \big|^2 \;
	{\rm Im}\big[\Pi(\nu, p,\Omega_p) \big] \;,
\end{equation}
where, from (\ref{eq:defpi}), one finds: 
\begin{equation}\label{eq:ImPi}
	{\rm Im}\big[\Pi(\nu, p,\Omega_p) \big] \;=\; \frac{\pi g^2}{ 2 p \Omega_p} \,
	\big[ \delta( \nu - p - \Omega_p) +  \delta( \nu + p + \Omega_p)
	\big] \;.
\end{equation}
Thus,
\begin{equation}\label{res:order1}
{\rm Im} [ \Gamma^{(1)}_p]\;=\; \frac{g^2}{8 \,\Omega_p} \,
	\int \frac{d^3p}{(2\pi)^3} \;\frac{1}{p} \, 
	\big| f({\mathbf p}, p + \Omega_p) \big|^2  \;.
\end{equation}

Since $p \geq 0$, note that, for this order to produce a non-vanishing
imaginary part, the frequency must overcome a threshold, namely, $|\nu| >
\Omega_p$. Of course, also $|\nu| > p$ must be satisfied. Those thresholds
may be identified as the frequencies for which the two $0+1$-dimensional
propagators involved in a 1-loop Feynman diagram become on-shell (one of
those propagators has a `mass' equal to $p$ and the other to $\Omega_p$). 
On physical grounds, the emission is produced when the center of mass motion is capable
of exciting the harmonic oscillator, and this happens only above the threshold.
As shown in Ref.\cite{PAMN}, the process involves the emission of single ``photons''
as opposed to the case of the usual DCE, in which there is pair creation.

Let us now consider the evaluation of $\Gamma_p^{(2)}[{\mathbf r}(t)]$.
\begin{align}\label{eq:gp_1}
\Gamma^{(2)}_p &=\; \frac{1}{4} \, \int
\frac{d^3{\mathbf p}}{(2\pi)^3}  
\int
\frac{d^3{\mathbf q}}{(2\pi)^3}  
\int\frac{dp^0}{2\pi} 
\int\frac{dq^0}{2\pi} \;
\int\frac{d\nu}{2\pi} \;
f({\mathbf p},p^0) 
f({\mathbf q},q^0 ) \nonumber\\ 
& \times f(-{\mathbf p}, -p^0 - \nu) f(-{\mathbf q}, -q^0 +\nu) 
\; C(p^0, q^0, \nu, {\mathbf p}, {\mathbf q}) \;,
\end{align}
with the kernel
\begin{align}\label{eq:defc}
 C(\omega, \nu, p^0, {\mathbf p}, {\mathbf q}) = - i \, g^4
\int\frac{d\omega}{2\pi} \Big[ 
\frac{1}{(\omega - \nu)^2 -\Omega_p^2 + i \epsilon} 
\frac{1}{\omega^2 -\Omega_p^2 + i \epsilon} 
\frac{1}{(\omega + p^0)^2 - {\mathbf p}^2 + i \epsilon} 
\frac{1}{(\omega - q^0)^2 - {\mathbf q}^2 + i \epsilon} 
\Big]\;.
\end{align}

Rather than writing the full expression for the imaginary part of
$\Gamma_p^{(2)}$, we consider now  its particular form, as well as for
$\Gamma_p^{(1)}$, for small amplitudes.  They may be  expanded in powers of
the departure of the particle from an equilibrium position ${\mathbf r}_0$.
Namely,  ${\mathbf y}(t)$, where ${\mathbf r}(t) = {\mathbf r}_0 +
\mathbf{y}(t)$.  

This requires to first expand: $f = f^{(0)} + f^{(1)} + f^{(2)} + \ldots$, where
$f^{(0)} = 2 \pi \, e^{- i {\mathbf p} \cdot {\mathbf r}_0} \delta(\nu)$ is
independent of the departure. In terms of $\tilde{y}^i(\nu)$, the
components of the
Fourier transform of ${\mathbf y}(t)$, the first and second order terms
in the expansion of $f$, are: 
\begin{align}
	& f^{(1)} \,=\,  - i \, e^{- i {\mathbf p} \cdot {\mathbf r}_0} \; p^i
	\tilde{y}^i(\nu) \;,\;\;
	f^{(2)} \,=\,  - \frac{1}{2} \, e^{- i {\mathbf p} \cdot {\mathbf r}_0}
	\; p^i p^j \, (\tilde{y}^i \star \tilde{y}^j)(\nu) \;, \nonumber\\
	& (\tilde{y}^i \star \tilde{y}^j)(\nu) \,=\, \int \frac{d\nu'}{2\pi} \; 
	\tilde{y}^i(\nu - \nu') \tilde{y}^j(\nu') \;.
\end{align}
Besides, we shall assume that ${\mathbf r}_0$ is the average position
around which the particle departs, so that $\tilde{y}^i(0) = 0$.

It is worth noting some general properties of the general terms in the
small-amplitude expansion of $f$. It is evident that higher order terms
involve higher convolution products of the Fourier transform of the
departure. That correspond to higher products of the departure itself.
Therefore, one sees that if the departure involves just one harmonic mode,
the $n$-order term will contain frequencies up to $n$-times the one of the
harmonic mode.


Then we see have for the first and second order terms in the expansion:

\subsection{First order effective action $\Gamma^{(1)}_p$}

For $\Gamma_p^{(1)}$,  up to the second order in ${\mathbf y}(t)$:
\begin{equation}
{\rm Im}[\Gamma^{(1)}_p] \;=\; \frac{1}{2} \,
\int  \,\frac{d\nu}{2\pi} \;  
	\vert \tilde{y}^j(\nu)\vert^2 \; m_p (\nu,\Omega_p), 
\end{equation} where 
\begin{equation}\label{mijno}
	m_p (\nu,\Omega_p) \;=\; \frac{g^2}{12 \pi \Omega_p} \,\delta^{ij} \, \theta(|\nu| -
	\Omega_p) \, \big(|\nu| - \Omega_p \big)^3 \;.
\end{equation}

\subsection{Second order effective action $\Gamma_p^{(2)}$}

Up to the second order in the amplitude, we also have
\begin{equation}\label{eq:Gamma2pert}
\Gamma^{(2)}_p \;=\; \frac{1}{2} \,\int \frac{d^3{\mathbf p}}{(2\pi)^3}  
\int \frac{d^3{\mathbf q}}{(2\pi)^3}  
\int \frac{d\nu}{2\pi}  
\; C(\nu,{\mathbf p}, {\mathbf q}) \; 
p^i p^j \; 
\tilde{y}^i(-\nu) \tilde{y}^j(\nu) 	
\;,
\end{equation}
with the kernel
\begin{equation}
 C(\nu, {\mathbf p}, {\mathbf q}) \,=\, - i \, g^4 
\int\frac{d\omega}{2\pi} \; 
	\frac{1}{(\omega - \nu)^2 - {\mathbf p}^2 + i \epsilon} \; 
	\frac{1}{\omega^2 - {\mathbf q}^2 + i \epsilon} \; 
	\frac{1}{[\omega^2 -\Omega_p^2 + i \epsilon]^2} \;.
\end{equation}

In order to evaluate this kernel, we write
\begin{equation}
\frac{1}{[\omega^2 -\Omega_p^2 + i \epsilon]^2} = \frac{d}{d\Omega_p^2}\frac{1}{[\omega^2 -\Omega_p^2 + i \epsilon]}\, ,
\end{equation}
decompose the last two factors in the integrand in partial fractions and use Eq.\eqref{eq:defpi1p}. The result is
\begin{equation}
	C(\nu,{\mathbf p}, {\mathbf q}) \; = \;  g^2 \frac{d}{d\Omega_p^2}\left[\frac{1}{q^2-\Omega_p^2}\left(\Pi(\nu,p,q)-\Pi(\nu,p,\Omega_p)\right)\right]\, .
\end{equation}	
Using Eq.\eqref{eq:ImPi}	we obtain
\begin{eqnarray}\label{eq:ImPif}
{\rm Im}\big[C(\nu, {\mathbf p}, {\mathbf q})\big]&=&\pi g^4\big[\frac{1}{pq}\frac{1}{(q^2-\Omega_p^2)^2}\delta(\nu-p-q) - \frac{1}{2 p\Omega_p^2}\frac{1}{q^2-\Omega_p^2}
\delta'(\nu-p-\Omega_p)\\ \nonumber
 &+&\frac{1}{2p\Omega_p^3}\frac{1}{(q^2-\Omega_p^2)^2} (q^2-3\Omega_p^2)\delta(\nu-p-\Omega_p)\big]\, ,
\end{eqnarray}
where we have taking into account that,  in order to obtain $\Gamma_P^{(2)}$, $C(\nu, {\mathbf p}, {\mathbf q})$ is multiplied 
by an even function of $\nu$.  Inserting this result into Eq.\eqref{eq:Gamma2pert} we obtain
\begin{equation}\label{ImGamma2pertfin}
{\rm Im}\big[\Gamma_p^{(2)}\big]=\frac{g^4}{24\pi^3}\int \frac{d\nu}{2 \pi}\vert \tilde y(\nu)\vert^2\Sigma(\nu, \Omega_p)\, ,
\end{equation}
where $\Sigma=\Sigma_1+\Sigma_2+\Sigma_3$ and
\begin{eqnarray}
\Sigma_1(\nu,\Omega_p)&=& \int_0^\nu dq\frac{q(\nu-q)^3}{(q^2-\Omega^2)^2}\\ \label{Sigma1}
\Sigma_2(\nu,\Omega_p)&=& \frac{3}{2}\theta(\nu-\Omega_p)\frac{(\nu-\Omega_p)^2}{\Omega_p^2}\int_0^\infty dq \frac{q^2}{(q^2-\Omega_p^2)}\\\label{Sigma2}
\Sigma_3(\nu,\Omega_p)&=&	\frac{1}{2}\theta(\nu-\Omega_p)\frac{(\nu-\Omega_p)^3}{\Omega_p^3}\int_0^\infty dq\frac{q^2(q^2-3\Omega_p^2)}{(q^2-\Omega_p^2)^2} 
\label{Sigma3}
\, .
\end{eqnarray}

Several comments are in order. We see that, in the second order, there is a
non vanishing contribution when the center of mass frequency is below the
threshold. This is the contribution coming from $\Sigma_1$ and is related
with the usual pair creation in the DCE, corrected here by the internal
structure of the moving particle. Indeed, $\Sigma_1$ comes from the term
proportional to $\delta(\nu-p-q)$ in Eq.\eqref{eq:ImPif}, that describes
the creation of a pair of particles with energies $p$ and $q$ respectively,
with the $\delta$ function
forcing energy conservation.

Above the threshold, the three terms contribute to the dissipative effects,
and constitute a correction to $\Gamma_p^{(1)}$. There are some subtle
points here. The integrals defining $\Sigma_i$ have potentials divergences
at $q=\Omega_p$ and for $q\to \infty$. One can readily check that the poles
at $q=\Omega_p$ do cancel when adding the three terms. However, $\Sigma_2$
and $\Sigma_3$ are linearly divergent in the ultraviolet, and thus
proportional to a $3-$momentum cutoff $\Lambda$. Due to the coupling to
the scalar field, the frequency $\Omega_p$ of the harmonic oscillator gets
renormalized with a divergent term proportional to $g^2\Lambda$ (see
Eq.\eqref{eq:renOmega}).
When working up to order $g^4$, this shift in the natural frequency must be
taken into account in the first order effective action. From Eq.\eqref{mijno} we obtain
\begin{equation}
m_p(\nu,\Omega_p)= 
m_p(\nu,\Omega_p^{ren})-\frac{g^4\Lambda}{48\pi^3(\Omega_p^{(ren)})^2}\left(\frac{(\nu-\Omega_p^{(ren)})^3}{\Omega_p^{(ren)}}+ 3(\nu-\Omega_p^{(ren)})^2\right)\, .
\end{equation}
It is easy to see that the extra terms in the above equation 
generate two extra terms in $\Gamma_p^{(1)}(\Omega_p)$, 
that cancel the divergences of $\Sigma_2$ and $\Sigma_3$. After this cancellation,  $\Gamma_p^{(2)}$ produces a finite correction to $\Gamma_p^{(1)}$. It is given by Eq.\eqref{ImGamma2pertfin} with $\Sigma\to\Sigma^{(ren)}$ and
\begin{eqnarray}
\Sigma_1^{\rm (ren)}(\nu,\Omega)&=& \int_0^\nu dq\frac{q(\nu-q)^3}{(q^2-\Omega^2)^2}\\ \label{Sigma1r}
\Sigma_2^{\rm (ren)}(\nu,\Omega)&=& \frac{3}{2}\theta(\nu-\Omega)\frac{(\nu-\Omega)^2}{\Omega_p^2}\int_0^\infty dq \left[\frac{q^2}{(q^2-\Omega^2)}-1\right]\\\label{Sigma2r}
\Sigma_3^{\rm (ren)}(\nu,\Omega)&=&	\frac{1}{2}\theta(\nu-\Omega)\frac{(\nu-\Omega)^3}{\Omega^3}\int_0^\infty dq\left[\frac{q^2(q^2-3\Omega^2)}{(q^2-\Omega^2)^2} -1\right]
\label{Sigma3r}
\, .
\end{eqnarray}

In order to simplify the notation we wrote $\Omega_p^{(ren)}\equiv \Omega$.  Although each $\Sigma_i$ has a pole at $q=\Omega$, the sum is finite.
Moreover, splitting the integrals as $\int_0^\infty=\int_0^\nu+\int_\nu^\infty$, $\Sigma^{\rm (ren)}$ can be computed analytically. We omit here the resulting 
long expressions, and plot $\Sigma^{\rm (ren)}/\Omega$   as a function of the dimensionless external frequency $\nu/\Omega$ in Fig.1. Below threshold, 
the result corresponds to the DCE due to the oscillation of the atom. In the limit $\nu\ll\Omega$ the result is proportional to $\nu^5$, which is
expected by dimensional analysis, since in this limit the effective coupling is $g^4/\Omega^4$ (see Eq.\eqref{eq:defc}). For $\nu\simeq\Omega$,
but still below threshold, the result includes the effect of the internal structure of the atom on the DCE.

\begin{figure}
	\includegraphics[scale=2]{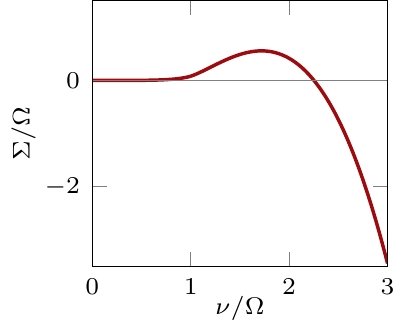}
	\label{fig:sigma}
	\caption{Second order correction to the imaginary part of the effective action for a particle in vacuum, whose internal degree of freedom is a QHO of frequency $\Omega$, and whose center-of-mass exhibits a single-frequency motion of frquency $\nu$.}
\end{figure}

Above threshold, the  second order result is a small  correction to the first order one, and combines both DCE and the emission of single photons
through excitation-deexcitation process. This correction to the imaginary part of the effective action goes as $- g^4(\nu/\Omega)^3$ for $\nu/\Omega\gg 1$ (note that the first order result is proportional to $g^2(\nu/\Omega)^3$ in this limit).


\section{Imperfect mirror: microscopic model}\label{sec:model mirror}

Dielectric slabs are in general nonlinear, inhomogeneous, dispersive and
also dissipative media. These aspects turn difficult the quantization of a
field when all of them have to be taken into account simultaneously. There
are different approaches to address this problem. On the one hand, one can
use a phenomenological description based on the macroscopic electromagnetic
properties of the materials. The quantization can be performed starting
from the macroscopic Maxwell equations, and including noise terms to
account for absorption. In this approach a canonical quantization scheme is
not possible, unless one couples the electromagnetic field to a reservoir,
following the standard route to include dissipation in simple quantum
mechanical systems. Another possibility is to establish a first-principles
model in which the slabs are described through their microscopic degrees of
freedom, which are coupled to the electromagnetic field. In this kind of
models, losses are also incorporated by considering a thermal bath, to
allow for the possibility of absorption of light. There is a large body of
literature on the quantization of the electromagnetic field in dielectrics.
Regarding microscopic models, the fully canonical quantization of the
electromagnetic field in dispersive and lossy dielectrics has been
performed by Huttner and Barnett (HB) \cite{hb92}. In the HB model, the
electromagnetic field is coupled to matter (the polarization field), and
the matter is coupled to a reservoir that is included into the model to
describe the losses. In the context of the theory of quantum open systems,
one can think the HB model as a composite system in which the relevant
degrees of freedom belong to two subsystems (the electromagnetic field and
the matter), and the matter degrees of freedom are in turn coupled to an
environment (the thermal reservoir). The indirect coupling between the
electromagnetic field and the thermal reservoir is responsible for the
losses.  It is well known that if we include the absorption, associated
with a dispersive medium, then the dielectric constant will be a complex
quantity, whose real and imaginary parts are related by the Kramers-Kronig
relations. Losses in quantum mechanics imply a coupling to a reservoir
whose degrees of freedom have to be added to the Lagrangian. This suggests
that, in order to quantize the vacuum field in a dielectric in a way that
is consistent with the Kramers-Kronig relations, one has to introduce the
medium into the formalism explicitly. This should be done in such a way
that the interaction between light and matter will generate both dispersion
and damping of the field.  The microscopic theory for the interaction
${\mathcal S}_I$ between the scalar field and
the imperfect mirror, consists of a tern of the form:
\begin{equation}
	{\mathcal S}_I^{(m)}(\phi) \;=\; -\frac{1}{2} \int_{x,y} \phi (x)
	V_{m}(x,y)
	\phi(y) \;.
\end{equation}
The kernel $V_m$, is a `potential' resulting from the integration of
microscopic degrees of freedom of the polarization field plus the external
reservoir. It can be regarded as a symmetric function of $x$ and $y$, since
the integrals in ${\mathcal S}_I$ symmetrize any bi-local function.

We then apply the above discussion to note that the potential $V_m$ originated by 
the interaction between the vacuum field and the imperfect mirror, may also be obtained by a
variant of the previous procedure for $V_p$; indeed, introducing a bosonic field
$Q(t,x^1,x^2)$, living on $x^3 = 0$, the plane occupied by the plate, playing the role of the polarization field, 
also couplet to an external (at equilibrium) environment with degrees of freedom denoted by $q_n(t,x^1,x^2)$, 
\begin{equation}\label{eq:gm}
e^{- \frac{i}{2} \int_{x,y} \phi(x) V_m(x,y) \phi(y) } 
 \;=\; 
\int {\mathcal D}Q \; e^{i {\mathcal S}^{eff}_m(Q,\phi)} \;.
\end{equation}
where the effective action ${\mathcal S}^{eff}_m(Q,\phi)$ is the result of integrating out the degrees of freedom $q_n$:

\begin{equation}\label{eq:gm2}
e^{i {\mathcal S}^{eff}_m(Q,\phi)} = \int {\mathcal D}q_n \; e^{i \left( {\mathcal S}_m(Q,\phi) + {\mathcal S}_m(Q,q_n)\right) } \;.
\end{equation}
where 
\begin{align}\label{eq:defsQ}
{\mathcal S}_m(Q,\phi) &=\;{\mathcal S}_m^{(0)}(Q) 
\,+\,{\mathcal S}_m^{int}(Q,\phi) \;, 
\nonumber\\
	{\mathcal S}^{(0)}_m(Q) &=\;\frac{1}{2}\int dt dx^1 dx^2\, 
	\big[ (\partial_t Q)^2 - ( \Omega_m^2 - i \epsilon) Q^2 \big] \;,
\nonumber\\
{\mathcal S}_m^{int}(q,\phi;{\mathbf r}) &=\; \gamma \,
	\int dt dx^1 dx^2 \,  Q(t,x^1,x^2) 
\; \phi(t,x^1,x^2,0) \;,
\end{align}
and 
\begin{align}\label{eq:defsq}
{\mathcal S}_m(q_n,\phi) &=\;{\mathcal S}_m^{(0)}(q_n) 
\,+\,{\mathcal S}_m^{int}(Q,q_n) \;, 
\nonumber\\
	{\mathcal S}^{(0)}_m(q_n) &=\;\frac{1}{2}\sum_n \int dt dx^1 dx^2\, 
	\big[ (\partial_t q_n)^2 - ( \omega_n^2 - i \epsilon) q_n^2 \big] \;,
\nonumber\\
{\mathcal S}_m^{int}(Q,q_n)) &=\; \sum_n \lambda_n \,
	\int dt dx^1 dx^2 \,  Q(t,x^1,x^2) 
\; q_n(t,x^1,x^2) \;.
\end{align}

Here, $\Omega_m$ is a frequency, and $\gamma$ determines
the coupling, which has the dimensions of $[{\rm mass}]^{3/2}$ (similar happens with $\omega_n$ and $\lambda_n$). 
Following~\cite{Farias:2014wca}, microscopic matter degrees of freedom on
the mirror, which we assume to occupy the $x^3 = 0$ plane, are assumed to
behave as one-dimensional harmonic oscillators, one at each point of the
plate.  Their generalized coordinates take values in an internal
space. Besides, no couplings between the coordinates at different points of the
plate are included, and there is a linear coupling between each oscillator and the
vacuum field and to the external reservoir. 

After integrating out the thermal environment, we obtain an effective action for 
the internal degrees of freedom into the mirror as

\begin{equation}
{\mathcal S}^{eff}_m(Q) = {\mathcal S}_m^{(0)}(Q)  + \int dt dx^1dx^2 Q(t, x^1,x^2) K(t,t') Q(t', x^1,x^2),
\end{equation}
where $K(t,t')$ is a nonlocal kernel that depends on the temperature and spectral density of the environment.
As is well known, for an environment formed by an infinite set of harmonic oscillators at high temperatures
with an ohmic spectral density this kernel becomes local and proportional to a dissipation coefficient $\xi$ \cite{QBM}.
In this limit, the effect of the environment on the in-out effective action can be taken into account just 
replacing $\Omega_m^2$ by $\Omega_m^2-i\xi$.

The interaction becomes then local, and $V_m$ has 
the form: 
\begin{equation}\label{eq:rvm}
	V_m(x,y) \;\equiv\; \gamma^2 \; \Delta_m(x^0-y^0) \, \delta({\mathbf
	x}-{\mathbf y}) \; \delta(x^3) \;,
\end{equation}
where $\gamma$ denotes a constant, which is the coupling between
the plate harmonic oscillator's degrees of freedom, and the vacuum field.
The precise form of $\Delta_m$ is more conveniently expressed in
terms of its Fourier transform, namely:
\begin{align}\label{eq:deflambdam}
\Delta_m(x^0-y^0) &=\; 
\int  \frac{d\nu}{2\pi} \, e^{-i \nu (x^0 - y^0)} \, 
\widetilde{\Delta}_m(\nu) \nonumber\\
\widetilde{\Delta}_m(\nu) &\equiv\; \frac{1}{\nu^2 - \Omega_m^2 + i
	\epsilon + i \xi} \; .
\end{align}

We would also be interested in the case of a mirror imposing
`perfect', i.e., Dirichlet, boundary conditions. Such a case
might be obtained by taking particular limits starting from a given $V_m$;
for example, 
\begin{equation}\label{eq:rvmd}
V_m(x,y) \;\to\; V_D(x,y) \,\equiv\, \eta \; \delta({\mathbf x_\parallel}-{\mathbf
	y_\parallel}) \, \delta(x^3) \, \delta(y^3) \;, \;\; \eta \to \infty \;,
\end{equation}
where we have adopted the notational convention that, for any spatial
vector ${\mathbf a}$,  ${\mathbf a_\parallel} \equiv (x^1, x^2)$. This
Dirichlet limit may be reached from different $V_m$ kernels, although it is
more convenient, given the simple geometry of the system
considered, to use images in order to write the exact scalar field
propagator. The same can be said about Neumann boundary conditions, for
which the field propagator in the presence of the mirror is also obtained
by using images.

\section{Moving atom in the presence of a plate} \label{sec:mirror}

In this Section, we deal with contributions which contain both $V_m$ and
$V_p$. We will work here, for the sake of simplicity, always to the first
order in $V_p$. Therefore, the structure of the term
calculated here is as follows:
\begin{equation}
\Gamma_{mp}[{\mathbf r}(t)] \;=\;- \frac{1}{2} \, \int_{x,y}  V_p(x,y)  \big\langle
\phi(x) \phi(y)\big\rangle_m \;,
\end{equation}
where now:
\begin{equation}
\langle \, \ldots \, \rangle_m \;\equiv\; 
	\frac{\int {\mathcal D}\phi \; \ldots \; e^{i [{\mathcal S}_0(\phi) 
	- \frac{1}{2} \int_{x,y} \phi(x) V_m(x,y) \phi(y)]}}{\int {\mathcal
	D}\phi \; e^{i [{\mathcal S}_0(\phi) - \frac{1}{2} \int_{x,y}
	\phi(x) V_m(x,y)\phi(y)]}} \;.
\end{equation}
In other words, the kind of contribution we consider here, looks like
$\Gamma_p^{(1)}$, albeit with the free propagator $\big\langle \phi(x)
\phi(y)\big\rangle_0$  replaced by the propagator in the presence of the
mirror,  $\big\langle \phi(x) \phi(y)\big\rangle_m$. The latter may be
incorporated either exactly or making some simplifying assumptions, in
order to be able to find the imaginary part in an explicit way.

The first non-trivial contribution, which we evaluate, arises when one
considers the expansion of the propagator to the first order in $V_m$, is:
\begin{equation}
\Gamma^{(1)}_{mp} \;=\; \frac{i}{2} \, \int_{x,y,x',y'}
\,  V_m (x,y) \,  V_p (x',y') \, G(x,x') \, G(y,y') \;,
\end{equation}
\begin{align}\label{eq:gint_2}
	\Gamma^{(1)}_{mp} \;=\; \frac{i}{2} \, \gamma^2 \, g^2 \int
	\frac{d^2{\mathbf p_\parallel}}{(2\pi)^2} & \int\frac{dp_3}{2\pi}
\int\frac{dq_3}{2\pi} \int\frac{d\omega}{2\pi}
	\int\frac{d\nu}{2\pi} \;\Big[ f({\mathbf p_\parallel},p^3,-\omega -
	\nu) \nonumber\\ 
\times f(-{\mathbf p_\parallel},q^3, \omega + \nu) \;
	&  \widetilde{\Delta}_m(\omega) \,  
\widetilde{G}(-\omega,{\mathbf p_\parallel},p^3)\, 
\,  \widetilde{\Delta}_p(\nu) \,  
\widetilde{G}(\omega,-{\mathbf p_\parallel},q^3) \Big]\, 
\end{align}
which, by a shift of variables may be written as follows:
\begin{equation}\label{eq:gint2_1}
	\Gamma^{(1)}_{mp} \;=\; \frac{1}{2} \, \int
	\frac{d^2{\mathbf p_\parallel}}{(2\pi)^2}  \int\frac{dp_3}{2\pi}
\int\frac{dq_3}{2\pi} \int\frac{d\nu}{2\pi} \;
f({\mathbf p_\parallel},p^3, -\nu) 
f(-{\mathbf p_\parallel},q^3,\nu) 
	\; B(\nu,{\mathbf p_\parallel},p^3,q^3)
\end{equation}
where
\begin{equation}\label{eq:defb}
B(\nu,{\mathbf p_\parallel},p^3,q^3) \;=\; i \,\gamma^2 \, g^2 \; 
	\int\frac{d\omega}{2\pi} \;  \widetilde{\Delta}_m(\omega) \,  
\widetilde{G}(-\omega,{\mathbf p_\parallel},p^3)\, 
\,  \widetilde{\Delta}_p(\nu - \omega) \,  
\widetilde{G}(\omega,-{\mathbf p_\parallel},q^3) \;. 
\end{equation}
Introducing the explicit form of the propagator in momentum space, and of
the $\widetilde{\Delta}$ functions, we see that:
\begin{align}\label{eq:defb1}
	B(\nu,{\mathbf p_\parallel},p^3,q^3) \,=\, - i \, \gamma^2 \, g^2 &
	\int\frac{d\omega}{2\pi} \Big[ \frac{1}{\omega^2 -\Omega_m^2 + i
	\epsilon + i \xi} \;  
	\frac{1}{\omega^2 - {\mathbf p_\parallel}^2 - (p^3)^2 + i \epsilon}
	\nonumber\\ 
	&\times \frac{1}{(\omega - \nu)^2 -\Omega_p^2 + i \epsilon} \;  
	\frac{1}{\omega^2 - {\mathbf p_\parallel}^2 - (q^3)^2 + i \epsilon} \Big]\;.
\end{align}
Since $B(\nu,{\mathbf p_\parallel},p^3,q^3) = B(\nu,{\mathbf
p_\parallel},q^3,p^3)$, we conclude that:
\begin{align}\label{eq:ig2_1}
	{\rm Im}\big[\Gamma^{(1)}_{mp}\big] &=\; \frac{1}{2} \, \int
	\frac{d^2{\mathbf p_\parallel}}{(2\pi)^2}  \int\frac{dp_3}{2\pi}
\int\frac{dq_3}{2\pi} \int\frac{d\nu}{2\pi} \;
f({\mathbf p_\parallel},p^3, -\nu) 
f(-{\mathbf p_\parallel},q^3,\nu) \nonumber\\
	&\times\; {\rm Im}\Big[B(\nu,{\mathbf p_\parallel},p^3,q^3) \Big] \;.
\end{align}

Now we come to the actual evaluation of $B$, which resembles a loop (box)
diagram on a $0+1$-dimensional quantum field theory. 

Introducing Feynman parameters, and integrating out $\omega$, after a
lengthy calculation we find:
\begin{align}\label{eq:B_1}
	B(\nu,{\mathbf p_\parallel},p^3,q^3)& =
	- \frac{\gamma^2 \, g^2}{2 \Omega_p}  
	\Big\{ \frac{\Omega_m +\Omega_p}{\Omega_m} \big[ \frac{1}{p^2 - \Omega_m^2 + i (\epsilon +
	\xi)} \frac{1}{q^2 - \Omega_m^2 + i (\epsilon + \xi)} \nonumber\\
	 & \hspace{2cm}\times \frac{1}{\nu^2 - (\Omega_m + \Omega_p)^2 + i
	(\epsilon+\xi)} \big]
	\nonumber\\
	&- \; \frac{1}{(p^2 - \Omega_m^2 + i \xi)(q^2 - p^2) p} \;\; 
\frac{p + \Omega_p}{\nu^2 - (p + \Omega_p)^2 + i \epsilon} \nonumber\\
	& - \; \frac{1}{(q^2 - \Omega_m^2 + i \xi)(p^2 - q^2)q} \;\; 
\frac{q + \Omega_p}{\nu^2 - (q + \Omega_p)^2 + i \epsilon} \Big\}\;.
\end{align}

Therefore, taking the imaginary part, the $\epsilon \to 0$ limit, and
keeping the leading terms when $\xi \to 0$,
\begin{align}\label{eq:B_3}
	{\rm Im}\big[B(\nu,{\mathbf p_\parallel},p^3,q^3)\big] &= \,
\frac{\pi \gamma^2 g^2 }{2 \Omega_p}  
\Big\{ \frac{\Omega_m + \Omega_p}{\Omega_m} \, 
 \delta_\xi\big(\nu^2 -(\Omega_m + \Omega_p)^2\big) \nonumber\\
	 \times \, \big[{\mathcal P}_\xi(p^2 - \Omega_m^2) \, {\mathcal
	 P}_\xi(q^2 -
	\Omega_m^2) \,- & \pi^2 \delta_\xi(p^2 - \Omega_m^2) \,
	\delta_\xi(q^2 - \Omega_m^2) \big]  \nonumber\\
-  \frac{p + \Omega_p}{p (q^2 - p^2)} \; {\mathcal P}_\xi(p^2 - \Omega_m^2) \; 
	\delta\big( \nu^2 - (p + \Omega_p)^2\big) & - \, \frac{q + \Omega_p}{q (p^2
- q^2)} \; {\mathcal P}_\xi(q^2 -
\Omega_m^2) \; \delta\big(\nu^2 - (q + \Omega_p)^2 \big) 
\Big\} \;,
\end{align}
where we have introduced notations for the approximants of Cauchy principal
value ${\mathcal P}$ and Dirac's $\delta$-function:
\begin{equation}\label{eq:approximants}
{\mathcal P}_\xi(x) \;=\; \frac{x}{x^2 + \xi^2} \;\; , \;\;\;
\delta_\xi(x) \;=\; \frac{1}{\pi} \, \frac{\xi}{x^2 + \xi^2} \;,
\end{equation}
respectively.

The terms retained in (\ref{eq:B_3}) above are meant to be the most
relevant, when $\xi \to 0$, regarding their contribution to the
integrals over frequency and momenta in the imaginary part of the effective
action. Besides, we have neglected terms which cancel each other in
${\rm Im}[B]$, in that limit.

On the other hand, note that, besides $\delta_\xi$, (\ref{eq:B_3}) also contains $\delta$
functions (they arise when taking the $\epsilon \to 0$ limit, and are
independent of $\xi$). Using standard $\delta$-function properties (note
that, in principle, they are not valid for $\delta_\xi$), we see that:
\begin{align}\label{eq:B_4}
{\rm Im}\big[B(\nu,{\mathbf p_\parallel},p^3,q^3)\big] \, =\,
\frac{\pi \gamma^2 g^2 }{2 \Omega_p}  
	\Big\{ & \frac{\Omega_m + \Omega_p}{\Omega_m} \, 
 \delta_\xi\big(\nu^2 -(\Omega_m + \Omega_p)^2\big) \,
 \big[{\mathcal P}_\xi(p^2 - \Omega_m^2) \, {\mathcal P}_\xi(q^2 -
\Omega_m^2) \nonumber\\
	& - \, \pi^2 \delta_\xi(p^2 - \Omega_m^2) \,
\delta_\xi(q^2 - \Omega_m^2) \big]  \nonumber\\
	\,- \, \frac{\theta(|\nu| - \Omega_p)}{2 (|\nu| - \Omega_p)} \, & {\mathcal
	P}_\xi\big((|\nu| - \Omega_p)^2 - \Omega_m^2\big) \;
	 \big[ \frac{\delta(p - |\nu| + \Omega_p)}{q^2 - (|\nu|-\Omega_p)^2 } +  
 \frac{\delta(q - |\nu| + \Omega_p)}{p^2 - (|\nu|-\Omega_p)^2 } \big]
\Big\}  \;.
\end{align}

We identify in (\ref{eq:B_4}) the sum of two contributions, with are quite
different regarding how and when they turn on, as functions of $\nu$.
Indeed, the first one has a $\delta_\xi$ function of $|\nu|$ minus the sum
of $\Omega_m$ and $\Omega_p$, while the second one contains a threshold at $|\nu|
= \Omega_p$. Also, for the latter to contribute, the function $f$ must be non-vanishing
when $|\nu|$ surpasses $p$ (or $q$). That will depend, of course, on the
nature of the motion considered.

In the following examples, depending on the nature of the motion involved
(reflected in $f$), we shall be able to consider the $\xi \to 0$ limit.
This will allow us to simplify the expressions as much as possible, namely,
depending on the smallest possible number of parameters. 

\subsection{Quantum friction}
The first example the we consider here corresponds to quantum friction,
namely, to motion with a constant velocity which is parallel to the plate:
\begin{equation}
	{\mathbf r}(t) \;=\; {\mathbf r}_0 + {\mathbf u} \, t  \;,
\end{equation}
with ${\mathbf u} = (u , 0, 0)$ and $ {\mathbf r}_0 = (0,0,a)$. We
have:
\begin{align}
	& f({\mathbf p_\parallel},p^3, \nu) \,=\, 2 \pi \,  e^{-i p^3 a} \,
\delta(\nu - p^1 u) \nonumber\\
	& f({\mathbf p_\parallel},p^3, - \nu) 
f(-{\mathbf p_\parallel},q^3,\nu ) \,=\, e^{-i
(p^3+q^3) a} \, 
T \, 2 \pi \, \delta(p^1 u + \nu) \;,
\end{align}
where $T$ denotes the extent of the time interval in the effective action
(which, for a constant velocity, must be extensive in time). We see that,
since $|u| < 1$,  $f$ will be non-vanishing only
when $|\nu| < p$, and therefore there will not be contributions to the
imaginary part coming from the term which has a threshold.
Therefore, we shall have:
\begin{align}\label{eq:igf}
	\frac{{\rm Im}[\Gamma^{(1)}_{mp}]}{T} &=\;(\Omega_m + \Omega_p) \,
	\frac{\pi \gamma^2 g^2 }{4 \Omega_p \Omega_m}  \int \frac{d^2{\mathbf
	p_\parallel}}{(2\pi)^2}  \int\frac{dp_3}{2\pi}
\int\frac{dq_3}{2\pi} \int\frac{d\nu}{2\pi} \;
e^{-i (p^3+q^3) a} \, 2 \pi \, \delta(p^1 u + \nu) \nonumber\\
& \, 
 \delta_\xi\big(\nu^2 -(\Omega_m + \Omega_p)^2\big) \,
 \big[{\mathcal P}_\xi(p^2 - \Omega_m^2) \, {\mathcal P}_\xi(q^2 -
\Omega_m^2) \, - \, \pi^2 \delta_\xi(p^2 - \Omega_m^2) \,
\delta_\xi(q^2 - \Omega_m^2) \big] 
  \;.
\end{align}
One sees first that the $\xi \to 0$ limit may be safely taken. Moreover,
the term which contains the product of three $\delta$-functions vanishes, 
since their simultaneous contribution requires a frequency which is larger
than $p$ or $q$ in modulus.  

Thus
\begin{equation}\label{eq:igf1}
\frac{{\rm Im}[\Gamma^{(1)}_{mp}]}{T} \;=\;
\frac{\pi \gamma^2 g^2 }{8\Omega_p \Omega_m}  \int \frac{d^2{\mathbf
	p_\parallel}}{(2\pi)^2}  \int\frac{dp_3}{2\pi}
	\int\frac{dq_3}{2\pi}  \; e^{-i (p^3+q^3) a}
	\frac{\delta\big(|p^1 u| -(\Omega_m + \Omega_p)\big)}{(p^2 -
	\Omega_m^2)\,(q^2 - \Omega_m^2)} \;,
\end{equation}
and, integrating out $p^3$ and $q^3$,
\begin{equation}\label{eq:igf_2}
	\frac{{\rm Im}\big[\Gamma^{(1)}_{mp}\big]}{T} \;=\; 
\frac{\pi \gamma^2 g^2}{32 \Omega_p \Omega_m} \, \int
\frac{d^2{\mathbf p_\parallel}}{(2\pi)^2} 
	\; \frac{e^{- 2 a\sqrt{ {\mathbf p_\parallel}^2 -
	\Omega_m^2}}}{{\mathbf p_\parallel}^2 - \Omega_m^2} 
	\delta\big( |p^1 u| - (\Omega_m +\Omega_p) \big) \;.
\end{equation}
We then make use of the remaining $\delta$ function to integrate out $p^1$:
\begin{equation}\label{eq:igf_3}
	\frac{{\rm Im}\big[\Gamma^{(1)}_{mp}\big]}{T} \;=\; 
	\frac{\gamma^2 \, g^2}{32 \pi \Omega_p \Omega_m u} \, 
	\int_0^\infty dp^2 
	\; \frac{e^{- 2 a\sqrt{ (p^2)^2  + \frac{1}{u^2}(\Omega_m +
	\Omega_p)^2 - \Omega_m^2}}}{(p^2)^2  + 
	\frac{1}{u^2}(\Omega_m + \Omega_p)^2 - \Omega_m^2} \;,
\end{equation}
or, 
\begin{equation}\label{eq:igf_4}
\frac{{\rm Im}\big[\Gamma^{(1)}_{mp}\big]}{T} \;=\; 
\frac{\gamma^2 \, g^2 \, a}{32 \pi \Omega_m \Omega_p} \,
\int_0^\infty dx \; \frac{e^{- \frac{2}{u}\sqrt{ x^2  + a^2 (\Omega_m +
\Omega_p)^2 - a^2 u^2 \Omega_m^2}}}{ x^2  + 
a^2(\Omega_m + \Omega_p)^2 - a^2 u^2 \Omega_m^2} \;,
\end{equation}
which has the proper dimensions and is consistent with previous results
corresponding to friction between planes; in this case the result becomes
proportional to the area of the planes, and the dimensionality of the
coupling $g$ is different (the same as that of $\gamma$).

\subsection{Small oscillations} 
In this example, we consider an expansion entirely analogous to the one of
the free oscillating particle, albeit now in the presence of the plate. We
then use the same expansion for the function $f$, namely,  $f = f^{(0)} + f^{(1)} +
f^{(2)} + \ldots$, with exactly the same terms. 
Inserting this expansion into the general expression for the imaginary part of $\Gamma_{mp}^{(1)}$,
and retaining up to terms of the second order in the departure ${\mathbf
y}$ from the equilibrium position, we see that the only surviving
contribution is the following:
\begin{equation}\label{eq:gosc_1}
	{\rm Im}[\Gamma^{(1)}_{mp}] =\frac{1}{2} \, \int
	\frac{d^2{\mathbf p_\parallel}}{(2\pi)^2}  \int\frac{dp_3}{2\pi}
\int\frac{dq_3}{2\pi} \int\frac{d\nu}{2\pi}
	f^{(1)}({\mathbf p_\parallel},p^3, -\nu) 
	f^{(1)}(-{\mathbf p_\parallel},q^3,\nu) 
	 {\rm Im}[B(\nu,{\mathbf p_\parallel},p^3,q^3)] \;.
\end{equation}
Inserting the explicit forms of (\ref{eq:B_4}) and  $f^{(1)}$ above, we note
that we may have contributions due to departures which are parallel or normal to
the plane will have a different weight; indeed, the remaining symmetries of
the system imply that the structure of the result is:
\begin{equation}\label{eq:gosc_2}
{\rm Im}[\Gamma^{(1)}_{mp}] \;=\;\frac{1}{2} \, \int\frac{d\nu}{2\pi}
	\big( \,m_\shortparallel(\nu) \, |\tilde{\mathbf
	y}_\shortparallel(\nu)|^2  \,+\, m_\perp(\nu) \, |\tilde{y}_3(\nu)
	|^2 \big) \;,
\end{equation}
depending on two scalar functions:
\begin{align}\label{eq:mpar}
m_\shortparallel(\nu) & = \, \frac{1}{2} \, 
\frac{\pi \gamma^2 g^2 }{2 \Omega_p}  
\Big\{ \frac{\Omega_m + \Omega_p}{\Omega_m} \, 
\delta_\xi\big(\nu^2 -(\Omega_m + \Omega_p)^2\big) \nonumber\\
\times \,\int \frac{d^2{\mathbf p_\parallel}}{(2\pi)^2} |{\mathbf p_\parallel}|^2 
	& \Big[  \big( \; \int \frac{dp^3}{2\pi}\, {\mathcal P}_\xi(p^2 - \Omega_m^2)
	e^{-i p^3 a} \;\big)^2  \, 
\, - \, \pi^2 \big( \; \int \frac{dp^3}{2\pi}\, \delta_\xi(p^2 - \Omega_m^2)
	e^{-ip^3 a} \;\big)^2 \Big]\nonumber\\
	\, - \, \frac{\theta(|\nu| - \Omega_p)}{(|\nu| - \Omega_p)} \,&
	{\mathcal P}_\xi\big((|\nu| - \Omega_p)^2 - \Omega_m^2\big) \;
	\int \frac{d^2{\mathbf p_\parallel}}{(2\pi)^2} |{\mathbf
	p_\parallel}|^2 \,
	\int \frac{dp^3}{2\pi} \int \frac{dq^3}{2\pi} 
	 \frac{\delta(p - |\nu| + \Omega_p)}{q^2 - (|\nu|-\Omega_p)^2} 
	e^{-i(p^3+q^3) a} \Big\}  \;,
\end{align}
and
\begin{align}\label{eq:mper}
m_\perp(\nu) & = \, - \frac{\pi \gamma^2 g^2 }{2 \Omega_p}  
\Big\{ \frac{\Omega_m + \Omega_p}{\Omega_m} \, 
\delta_\xi\big(\nu^2 -(\Omega_m + \Omega_p)^2\big) \nonumber\\
\times \,\int \frac{d^2{\mathbf p_\parallel}}{(2\pi)^2} 
	& \Big[  \big( \; \int \frac{dp^3}{2\pi}\, p^3 \, {\mathcal P}_\xi(p^2 - \Omega_m^2)
	e^{-i p^3 a} \;\big)^2  \, 
\, - \, \pi^2 \big( \; \int \frac{dp^3}{2\pi}\,p^3 \,  \delta_\xi(p^2 - \Omega_m^2)
	e^{-ip^3 a} \;\big)^2 \Big]\nonumber\\
	\, - \, \frac{\theta(|\nu| - \Omega_p)}{(|\nu| - \Omega_p)} \,&
	{\mathcal P}_\xi\big((|\nu| - \Omega_p)^2 - \Omega_m^2\big) \;
	\int \frac{d^2{\mathbf p_\parallel}}{(2\pi)^2} 
	\int \frac{dp^3}{2\pi} \int \frac{dq^3}{2\pi} 
p^3 \, q^3 \, \frac{\delta(p - |\nu| + \Omega_p)}{q^2 - (|\nu|-\Omega_p)^2} 
	e^{-i(p^3+q^3) a}\, \Big\}  \;.
\end{align}

\begin{figure}
	\includegraphics[scale=1.5]{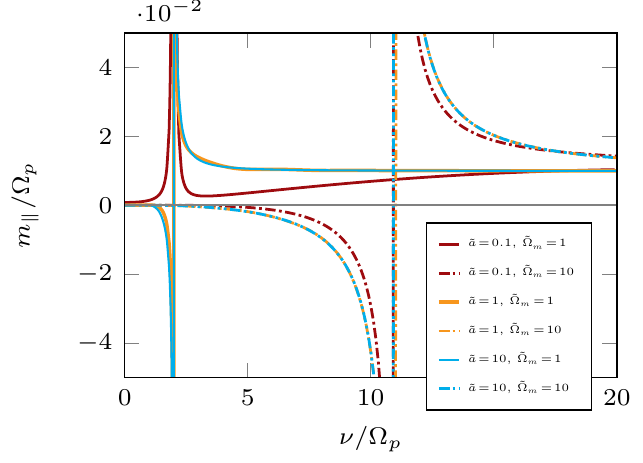}
	\caption{\label{fig:m_parallel}Second order correction to the imaginary part of the effective action, for a particle, modeled as a QHO of frequency $\Omega_p$, whose center-of-mass moves with small oscillations of frequency $\nu$, parallel to the plane, at a distance $a$ above it. The plane is modeled as a continuous set of harmonic oscillators of frequency $\Omega_m$. We have defined $\tilde{a}=a \Omega_p$, $\tilde{\Omega}_m=\Omega_m / \Omega_p$, and we have set the dissipation of the plate as $\xi / \Omega_p^2 = 0.01$.}
\end{figure}

After performing the integrals over $p^3$ and $q^3$, we find for
$m_\shortparallel$ a more explicit expression:
\begin{align}\label{eq:mpar1}
	m_\shortparallel(\nu) & = \, \frac{1}{2} \, 
\frac{\pi \gamma^2 g^2 }{2 \Omega_p} 
\Big\{ \frac{\Omega_m + \Omega_p}{4 \pi \Omega_m} \, 
\delta_\xi\big(\nu^2 -(\Omega_m + \Omega_p)^2\big) \, A_\shortparallel(\xi,\Omega_m,a)		\nonumber\\
	& + \, \frac{1}{8 \pi^2} \,  \theta(|\nu| - \Omega_p) \, (|\nu| - \Omega_p)^2  \,
{\mathcal P}_\xi\big((|\nu| - \Omega_p)^2 - \Omega_m^2\big) \; B_\shortparallel((|\nu| - \Omega_p) a ) 	 \Big\} \;,
\end{align}
with
\begin{eqnarray}
&& A_\shortparallel(\xi,\Omega_m,a)=\int_{-\Omega_m^2}^\infty \frac{du\, u }{u^2 +\xi^2} \,
e^{-2 \beta a} \, [ u \, \cos( 2 \alpha a) + \xi \, \sin( 2 \alpha a) ]\, ,\nonumber\\
&& B_\shortparallel((|\nu| - \Omega_p) a ) = \int_0^1 du \,[\frac{1-u^2}{u} \, \sin(2(|\nu| - \Omega_p) a u)]\, ,
\end{eqnarray}
and
\begin{equation}\label{eq:defab}
\alpha \equiv \sqrt{\frac{\sqrt{u^2 + \xi^2} - u}{2}} \;\;,\;\;\;
\beta \equiv \sqrt{\frac{\sqrt{u^2 + \xi^2} + u}{2}} \;.
\end{equation}	
In obtaining (\ref{eq:mpar1}), no small-$\xi$ approximation has been made, and the results are shown in Fig. \ref{fig:m_parallel} for different values of the parameters of the material (distance to the plate $a$ and frequency $\Omega_m$). The plots show a resonant behaviour for $|\nu | = \Omega_m + \Omega_p$, and the specific shape of this resonance depends on the distance $a$, as we will discuss below.

\begin{figure}
	\includegraphics[scale=1.5]{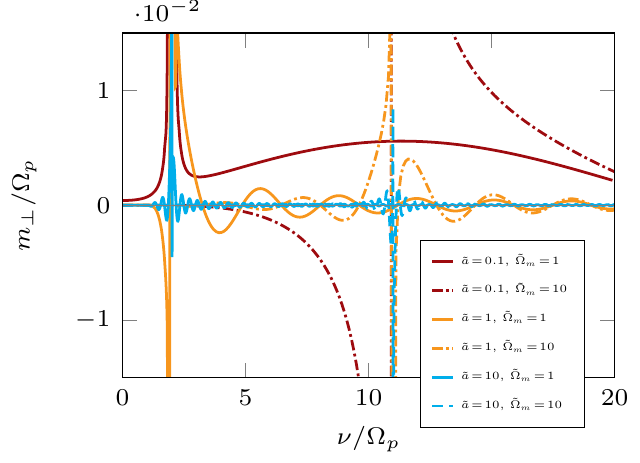}
	\caption{\label{fig:m_perp}Second order correction to the imaginary part of the effective action, for a particle, modeled as a QHO of frequency $\Omega_p$, whose center-of-mass moves with small oscillations of frequency $\nu$ at a distance $a$, normal to a plane. The plane is modeled as a continuous set of harmonic oscillators of frequency $\Omega_m$. We have defined $\tilde{a}=a \Omega_p$, $\tilde{\Omega}_m=\Omega_m / \Omega_p$, and we have set the dissipation of the plate as $\xi / \Omega_p^2 = 0.01$.}
\end{figure}

An entirely similar procedure allows us to find:
\begin{align}\label{eq:mperp1}
	m_\perp(\nu) & = \, 
\frac{\pi \gamma^2 g^2 }{2 \Omega_p} 
\Big\{ \frac{\Omega_m + \Omega_p}{16  \pi \Omega_m} \, 
\delta_\xi\big(\nu^2 -(\Omega_m + \Omega_p)^2\big) \,
A_\perp(\xi,\Omega_m,a)	
	\nonumber\\
	& - \, \frac{1}{8 \pi^2} \,  \theta(|\nu| - \Omega_p) \, 
	(|\nu| - \Omega_p)^2  \,
{\mathcal P}_\xi\big((|\nu| - \Omega_p)^2 - \Omega_m^2\big) \;
	 B_\perp((|\nu| - \Omega_p) a ) \Big\}\;,
\end{align}
with $\alpha$ and $\beta$ as in (\ref{eq:defab}) and
\begin{eqnarray}
&& A_\perp(\xi,\Omega_m,a)=\int_{-\Omega_m^2}^\infty \, du\, 
e^{-2 \beta a} \, \cos( 2 \alpha a)  \, ,\nonumber\\
&& B_\perp((|\nu| - \Omega_p) a )=\int_0^1 du \, u \, \sin(2(|\nu| - \Omega_p) a u)\, .
\end{eqnarray}
We show $m_\perp$ in Fig. \ref{fig:m_perp} for different characteristics of the materials. The same resonant behaviour is observed near $|\nu| = \Omega_p + \Omega_m$, and also an oscillatory behaviour that was absent for $m_\parallel$. These oscillations have a frequency that depends on the distance to the plate, and their presence is related to the fact that that distance is indeed modified by the center-of-mass motion, in contrast to what happens to the parallel contribution.

The coefficients $B_\shortparallel$ and $B_\perp$ can be computed
explicitly, the result being
\begin{eqnarray}
&& B_\shortparallel(x)= -B_\perp(x) + Si(2x)\, ,\nonumber\\
&& B_\perp(x)= \frac{-2 x \cos(2x)+\sin(2x)}{4 x^2}\, ,
\end{eqnarray}
where $Si(x)$ is the sine-integral function.

It is worth noting that there is a qualitative difference in the $a \to
\infty$ behaviours of $m_\shortparallel$ and $m_\perp$ above the $|\nu| >
\Omega_p$ threshold. Indeed, while the latter vanishes, the former reaches
the finite limit:
\begin{equation}\label{eq:mpar2}
m_\shortparallel(\nu) \; \to  \; \frac{\gamma^2 g^2 }{64\Omega_p} 
\, (|\nu| - \Omega_p)^2 \, {\mathcal P}_\xi\big((|\nu| - \Omega_p)^2 -
\Omega_m^2\big) \;.
\end{equation}
The difference between the response for the two different kinds of
oscillations can be traced back to the fact that the respective effective
actions depend on different properties of the scalar field propagator
in the presence of the plate. Indeed, for parallel motion, one needs the
propagator between two points at the same distance $a$ from the plate,
while for perpendicular motion one has to take two derivatives with respect
the to the third coordinate, and then to evaluate at the average distance, $a$. 
This results in different $a \to \infty$ limits. Physically, this may be
interpreted as a consequence of the different response properties of the
plate for normal vs parallel incidence.

\begin{figure}
	\includegraphics[scale=1.5]{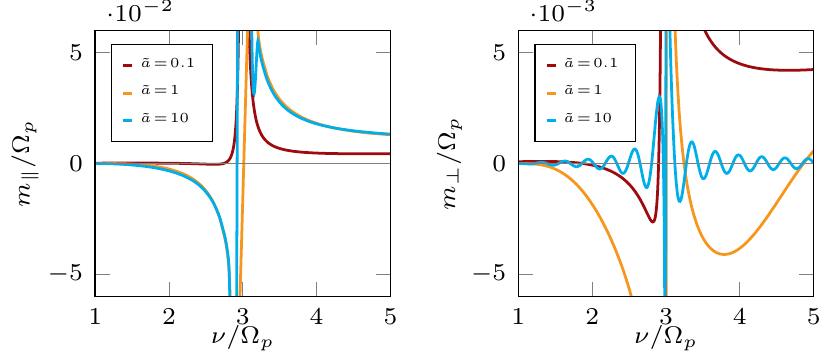}
	\caption{\label{fig:plots_m}Second order correction to the imaginary part of the effective action for $\Omega_m / \Omega_p=2$. We show the behaviour near the resonance, that occurs at $|\nu| = \Omega_m + \Omega_p$. We have defined $\tilde{a}= a \Omega_p$.}
\end{figure}

We see that, up to this order, the emission probability for both parallel
and perpendicular motions is a combination of the approximants of  Cauchy's
principal value ${\mathcal P}$ and Dirac's $\delta$-function (see
Eq.\eqref{eq:approximants}), both localized at the resonant frequency
$\vert\nu\vert =  \Omega_m + \Omega_p$. Moreover, in the limit $\xi\to 0$
the coefficients are  finite and can be computed explicitly; using the fact
that $\alpha\simeq 0$ and $\beta\simeq \sqrt{\vert u\vert}$.  We obtain:
\begin{eqnarray}
A_\shortparallel(0,\Omega_m,a)&=& A_\perp(0,\Omega_m,a) = \int_{-\Omega_m^2}^\infty \, du\, 
e^{-2 \sqrt{\vert u\vert}a}\nonumber\\
&=& \frac{2}{a^2}\big(2-(1+\Omega_m a)e^{-\Omega_m a}\big)\, .
\end{eqnarray}
It is interesting to remark that the emission probabilities have  peaks at the resonant frequency (with a height determined by he coefficients of the $\delta_\xi$ functions) and regions of enhancement and suppression at both sides of the resonant frequency, with amplitudes given by the coefficients that multiply the principal values. The ratio of the coefficients of $\delta_\xi$
and ${\mathcal P}_\xi$ in Eqs.\eqref{eq:mpar1} and \eqref{eq:mperp1} determine, for each kind of motion, which is the dominant behaviour. We illustrate this in Fig. \ref{fig:plots_m}. 

The structure of the results is reminiscent to what happens with the spontaneous decay rate of an excited atom immersed in an absorbing dielectric \cite{spontaneous}.  Note also that the dissipation coefficient $\xi$ regulates the otherwise
infinite results that would be obtained for non-lossy dielectrics. In the electromagnetic case, the presence of $\xi$ insures the validity of the Kramers-Kronig relation.  For the particular case ($\xi=0$), one should work beyond the perturbative approximation in the interaction between the mirrors' degrees of freedom and the quantum field.  In our case, the atom is outside the dielectric, and the corrections to the free space probability of emission are due to the vacuum fluctuations that are present near the surface of the dielectric plane \cite{Bartolo}.

\section{Conclusions}\label{sec:conc}
We have calculated the vacuum persistence amplitude for a moving harmonic
oscillator, first in free space,  and afterwards in the presence of a
dielectric plane. The in-out effective action was
perturbatively evaluated, in an expansion in powers of the coupling between
the atom and the field. We presented the result for the corresponding
imaginary part as a functional of the atom's trajectory, showing that,   
to the lowest non trivial order, there is a threshold.
This is associated to the possibility of internal excitation of the atom,
before radiation emission. We also found that the NTLO exhibits the combination
of the previously mentioned effect with the usual DCE (which does not 
involve such excitation process).
An interesting  point of the calculation is the shift in the natural
frequency of the oscillator (and therefore in the energy levels of the
atom) produced by the vacuum fluctuations. It is mandatory to take into
account this shift in order to obtain finite corrections to the vacuum
persistence amplitude at the NTLO.  Further, we have considered the motion of
the atom in the presence of an imperfect mirror, considering quantum
harmonic oscillators as microscopic degrees of freedom coupled to an
environment as a source of internal dissipation. Again, we have evaluated
the vacuum persistence amplitude for the case of an atom moving near the
plate, up to first order in both couplings between the atom and the
microscopic degrees of freedom and the vacuum field. We have shown that,  at
the same order in which there is DCE, there also is quantum contactless friction
and corrections to free emission. These corrections show a peculiar
behaviour when the external frequency equals the sum of the frequency of
the atom and the frequency of the microscopic degrees of freedom, with
regions of enhancement and suppression of the vacuum persistence amplitude.
We pointed out that this is similar to what happens with the spontaneous
emission of an atom immersed into a lossy dielectric. The inclusion of
losses in the dielectric is crucial to get a finite vacuum persistence
amplitude for an accelerated motion of the atom. Friction effects are less
sensitive to dissipation, and have a well defined limit for non-lossy
dielectrics.

\section{Acknoweledgements}

This work was supported by ANPCyT, CONICET, UBA and UNCuyo; Argentina. M. B. Far\'\i as acknowledges financial support from the national Research Fund Luxembourg under CORE Grant No. 11352881.

\end{document}